\RequirePackage{fix-cm}
\documentclass[twocolumn]{svjour3}          
\smartqed  
\usepackage{graphicx}
\usepackage{times,mathptmx}  
\usepackage{hyperref,amsmath,amssymb,color}
\usepackage[misc,geometry]{ifsym}
\journalname{Granular Matter}
\begin{document}

\title{Shear dispersion in dense granular flows
}


\author{Ivan C. Christov \and Howard A. Stone}


\institute{I. C. Christov (\Letter) and H. A. Stone \at
           Department of Mechanical and Aerospace Engineering, Princeton University, Princeton, NJ 08544, USA\\
             \email{christov@alum.mit.edu}\\
             \emph{Present address} of I. C. Christov: Theoretical Division and Center for Nonlinear Studies, Los Alamos National Laboratory, Los Alamos, NM 87545, USA   
}

\date{Received: date / Accepted: date}

\maketitle

\begin{abstract}
We formulate and solve a model problem of dispersion of dense granular materials in rapid shear flow down an incline. The effective dispersivity of the depth-averaged concentration of the dispersing powder is shown to vary as the P\'eclet number squared, as in classical Taylor--Aris dispersion of molecular solutes. An extensions to generic shear profiles is presented, and possible applications to industrial and geological granular flows are noted.%
\end{abstract}

\section{Introduction}
\label{intro}

Dispersal of a passive solute, such as a dye in a pipe flow or a pollutant in a river, is a classical fluid mechanics transport phenomenon that falls within the subject of macrotransport processes \cite{be93}. G.\ I.\ Taylor \cite{t53}, followed by Aris \cite{a56}, showed that the dispersal of a passive solute in a pressure-driven laminar flow in a circular pipe of radius $R$ can be described, at long times and far downstream from its injection point, by a cross-sectionally averaged advection-diffusion process in which the mean solute concentration $\bar{c}$ is advected by the mean flow $\overline{v_x}$ but diffuses with an \emph{effective dispersivity} $\mathcal{D}$ that depends on its molecular diffusivity $D_\mathrm{m}$, the mean flow speed $\overline{v_x}$, and the typical length scale $R$ associated with the cross-section of the flow vessel. In particular, $\mathcal{D} = D_\mathrm{m} + \overline{v_x}^2 R^2/(48 D_\mathrm{m})$ \cite{be93,t53,a56}. (Note that $\mathcal{D}$ is undefined in the limit of $D_\mathrm{m}\to0$ because non-diffusive solutes are simply advected by the flow and remain on the streamlines they start on for all time.) Many variations of the classical Taylor dispersion problem have been considered in the fluid mechanics literature \cite{be93,yj91}. Although the phenomenon has been mentioned in studies of self-diffusion of granular materials in shear flow, in which the diffusivity is inferred from the mean squared displacement \cite{nht95,c97}, to the best of our knowledge the dispersion problem has not been posed ``in the spirit of Taylor'' for rapidly flowing dense granular materials, despite the fact that the latter can behave similar to fluids and can be approximated as a continuum \cite{s84,jnb96,afp13,at09}. 

At the same time, there are practical implications to understanding the spread and dispersal of one type of granular material, such as a pharmaceutical powder, glass beads in the laboratory, or rocks and vegetation in a landslide, in a second granular material. For example, understanding granular dispersion is relevant for industrial separation processes such as the drying of powders for the purposes of dehydrating food \cite{hd08}. Another aspect to this process is the vibration of the vessel with the goal of mixing a flowing powder with another powder injected into the flow via diffusion in the transverse direction \cite{setal08}. 

Modeling transport of particulate materials is also important in geophysical flows such as snow avalanches, mud and land slides \cite{i97,ph07}. For example, in a polydisperse ava\-lanche, segregation drives the large particles to the front \cite{gk10}, which can lead to fingering instabilities \cite{pds97}. The resulting distribution of debris upon the cessation of flow can dictate the ecological impact of the event \cite{nist93}. Hence, it is important to know how the various constituent materials are dispersed during the landslide. More quantitatively, we can estimate the relevance of shear dispersion in the geophysical context by noting that a typical landslide can reach speeds up to $\overline{v_x} \simeq 10$ m/s, has a runout distance $\ell \simeq 10-100$ km, a depth of $h \simeq 0.5-1$ m, and an effective diameter $d\simeq 1$ mm$-1$ m for the particulate material \cite{i97}. Let us estimate the debris as being relatively fine, $d \simeq 10$ cm, and thus more likely to be monodisperse. Then, the diffusivity can be estimated by dimensional considerations as $D_0 \propto d^2 \overline{v_x}/h \simeq 10^{-1}$ m$^2$/s (see the discussion in Section~\ref{sec:flow_incline} below), from which we estimate, based on an analogy to Taylor's result \cite{t53}, the shear-augmented portion of the effective dispersivity as $\overline{v_x}^2h^2/D_0 \simeq 10^3$ m$^2$/s. For a laboratory-scale chute flow experiment, on the other hand, the typical values are $D_0 \simeq 10^{-6}$ m$^2$/s, $\overline{v_x} \simeq 1$ m/s and $h \simeq 10^{-2}$ m \cite{hh80}, which gives $\overline{v_x}^2h^2/D_0 \simeq 10^{-2}$ m$^2$/s. Both of these estimates indicate that the shear-augmented portion of the effective dispersivity is not negligible, specifically it is several orders of magnitude larger than $D_0$.

Thus, the goal of the present work is to pose the shear dispersion problem for rapid flows of particulate materials and to present solutions for the effective dispersivity for some elementary dense granular flows. We restrict our discussion to dry, cohesionless monodisperse materials to avoid, in particular, the complicating effects of segregation of bidisperse and polydisperse mixtures due to flow \cite{sl88}. By ``solute'' we mean a set of tagged particles released at the upstream end of the flow ($x=0$ in Fig.~1 below).


\section{Mathematical theory of shear dispersion}
\label{sec:math_disp}

Consider a steady two-dimensional (2D) flow $v_x(z)$ that is uniform in $x$ with $x\in[0,\infty)$ as the streamwise coordinate and $z\in[0,h]$ as the transverse coordinate. The evolution of the concentration $c$ (number of particles per unit area) of a diffusive passive tracer with (non-constant) diffusivity $D$ advected by such a flow obeys
\begin{equation}
\frac{\partial c}{\partial t} + v_x(z)\frac{\partial c}{\partial x} = \frac{\partial}{\partial x}\left(D\frac{\partial c}{\partial x}\right) + \frac{\partial}{\partial z}\left(D\frac{\partial c}{\partial z}\right).
\label{eq:adv_diff}
\end{equation}
Equation~\eqref{eq:adv_diff} is supplemented with no-flux boundary conditions $\partial c/\partial z = 0$ at $z = 0,h$, since material is not allowed to leave through the layer's boundaries, an initial condition $c(x,z,0)=c_i(x,z)$, and decay boundary conditions $c\to 0$ as $|x|\to\infty$.

Formally, we can always let $c(x,z,t) \equiv \bar{c}(x,t) + c'(x,z,t)$ and $v_x(z) \equiv \overline{v_x} + v_x'(z)$, where an overline denotes the depth-averaging operator $\overline{(\cdot)} = \frac{1}{h}\int_0^h (\cdot) \,\mathrm{d}z$, and primes denote deviation from the average. By construction, the overlined quantities can only depend on the axial coordinate $x$ and time $t$ and $\overline{c'} = \overline{v_x'} = 0$. Then, following Taylor \cite{t53}, we analyze the flow in the limit that the transverse diffusion time $h^2/D_0$ is much shorter than the typical streamwise advection time $\ell/\overline{v_x}$, where $\ell$ is a characteristic axial length scale over which we study the flow, and $D_0$ is a characteristic diffusivity. Based on the estimates given in the introduction, $h^2/D_0\simeq 10$ s and $\ell/\overline{v_x}\simeq 10^3-10^4$ s for a geophysical debris flow.\footnote{For the laboratory-scale chute flow from \cite{hh80}, $\ell \simeq 1$ m, so $h^2/D_0\simeq 10^2$ s and $\ell/\overline{v_x}\simeq 1$ s. In this particular experimental setup, we would not expect to see dispersion because the granular layer is too thin, and the device is too short in the streamwise direction.} Therefore, $\ell/h\gg \overline{v_x}h/D_0$ and, for $|c'|/\bar{c} \ll 1$, the evolution of the mean $\bar{c}$ separates from the fluctuations $c'$, leading to a one-way coupled set of macrotransport equations. In general, for $D = D(c,x,z,t)$, one obtains an advection-diffusion equation for the mean concentration $\bar{c}$ and an ordinary differential equation for the spatial structure of the fluctuations (see the appendix):
\begin{align}
\frac{\partial \bar{c}}{\partial t} + \overline{v_x}\frac{\partial \bar{c}}{\partial x} &\approx \frac{\partial}{\partial x}\left(\overline{D}\frac{\partial \bar{c}}{\partial x}\right) - \overline{v_x'\frac{\partial c'}{\partial x}},\label{eq:cbar}\\
\frac{\partial}{\partial z}\left({D}\frac{\partial c'}{\partial z}\right) &\approx v_x'\frac{\partial \bar{c}}{\partial x},
\label{eq:cprime}
\end{align}
where $\overline{D}$ is the depth-averaged diffusivity.

Equation \eqref{eq:cprime} can be integrated, and then the fluctuation induced diffusive flux, i.e., the last term on the right-hand side of Eq.~\eqref{eq:cbar}, can be evaluated using the fact that $v'_x$ is independent of $x$:%
\begin{equation}
\overline{v_x'\frac{\partial c'}{\partial x}} = \frac{\partial}{\partial x}\left[ \overline{ v_x'(z)\int_0^z \frac{1}{D} \int_0^{\tilde{z}} v_x'(\tilde{\tilde{z}}) \,\mathrm{d}\tilde{\tilde{z}} \,\mathrm{d}\tilde{z}} \frac{\partial \bar{c}}{\partial x}\right].
\label{eq:fluc_flux}
\end{equation}
Combining Eqs.~\eqref{eq:cbar} and \eqref{eq:fluc_flux}, we can define the effective dispersivity of $\bar{c}$ (see also \cite{be93,gs12}) as
\begin{equation}
\mathcal{D} = \frac{1}{h} \int_0^h D \,\mathrm{d}z - \frac{1}{h} \int_0^h v_x'(z)\int_0^z \frac{1}{D} \int_0^{\tilde{z}} v_x'(\tilde{\tilde{z}}) \,\mathrm{d}\tilde{\tilde{z}} \,\mathrm{d}\tilde{z} \,\mathrm{d}z.
\label{eq:d_eff}
\end{equation}
The first term is the influence of the basic diffusion process alone, while the second terms gives the contribution of the shear via the ``fluctuations'' $v_x'$ in the velocity.


\section{Rapid granular flow down an inclined plane}
\label{sec:flow_incline}

\begin{figure}
	\centerline{\includegraphics[width=0.825\columnwidth]{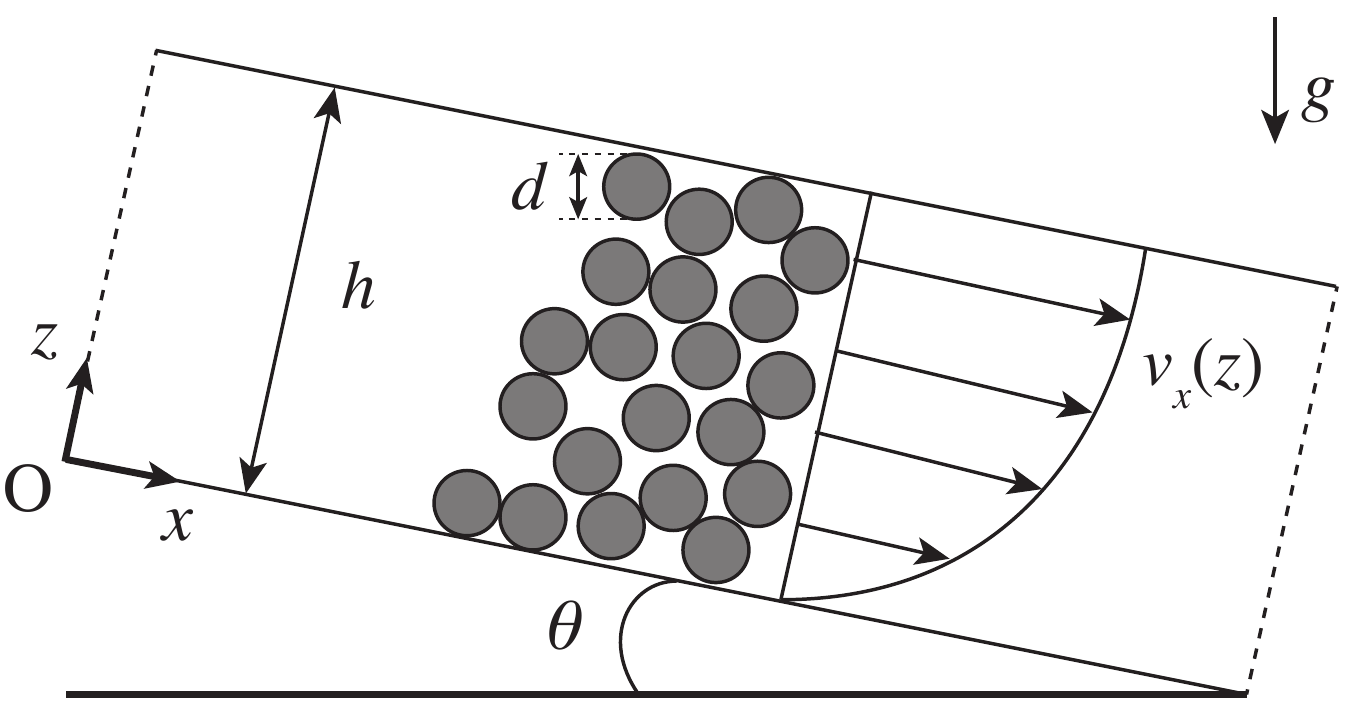}}
	\caption{Schematic of a rapid dense granular shear flow down an incline at an angle $\theta$. The granular material is assumed to be dry, cohesionless and monodisperse (i.e., the particles are of identical size, density, surface roughness, etc.) and the flow is steady and fully developed so that it can be approximated by the continuous profile $v_x(z)$ at any streamwise location $x$. The layer is typically dozens to hundreds of particles thick, hence $d/h\ll1$.}
\label{fig:shear_flow}	
\end{figure}

Consider the flow of a granular material down an incline at an angle $\theta$ with respect to the horizontal, as shown in Fig.~\ref{fig:shear_flow}. We assume the flow is fully developed and steady, and the thickness of the layer is approximately $h$ everywhere. The local viscoplastic rheology model \cite{jfp06} can be used to show \cite{afp13,at09,k11} that the local shear rate varies as the square root of the local depth:
\begin{equation}
	\dot\gamma \equiv \frac{\partial v_x}{\partial z} = A\sqrt{h-z},
\label{eq:const_rel}
\end{equation}
where $A$ is a constant. Typically, this type of model corresponds to an experiment performed at constant pressure at the free surface, so that the pressure distribution throughout the layer is hydrostatic \cite{afp13}. Under these conditions, the layer thickness $h$ can fluctuate.\footnote{Streamwise variations of the layer thickness of the form $h(z) = h_0[1+\beta f(z)]$ have been shown to lead to contributions on the order of $\beta^2$ to the effective dispersivity $\mathcal{D}$ \cite{bdlb09}. Hence, streamwise variations of the layer could be incorporated into the dispersion calculation, by replacing $h$ with $h(z)$ everywhere, without changing the result,  as long as the variations are small, i.e., $\beta = \mathcal{O}(h_0/\ell) \ll 1$, which renders the $\mathcal{O}(h_0^2/\ell^2)$ contributions to $\mathcal{D}$ negligible within the chosen order of approximation (see the appendix). Furthermore, we expect that the Bagnold profile remains valid for such $h(z)$ with $\beta\ll1$.}  However, here, we assume $h\approx const.$ and similarly the volume fraction $\phi\approx const.$ to a first approximation. This assumption is consistent with experiments \cite{afp13}. Thus, $h$ is representative of the thickness of the layer of \emph{fluidized} material, not of the static packing prior to flow.

Integrating Eq.~\eqref{eq:const_rel} and enforcing ``no slip'' at the bottom surface, $v_x(0) = 0$, yields the classical Bagnold profile \cite{b54,seghlp01}:
\begin{multline}
	v_x(z) = \frac{2}{3}A\left[h^{3/2} - \left(h-z\right)^{3/2}\right],\\ A = \frac{I_0}{d}\left(\frac{\tan\theta - \tan\theta_0}{\tan\theta_2 - \tan\theta}\right)\sqrt{\phi g\cos\theta},
\label{eq:Bagnold}
\end{multline}
where $d$ is the particle diameter, $I_0$ is a dimensionless model parameter, $\theta_0$ is the marginal angle of repose at which flow begins, $\theta_2$ is the angle beyond which steady flow is impossible, $\phi$ is the volume fraction,\footnote{That is, the proportion of volume occupied by the number of particles in a unit area. Note that $c$ is the concentration of the injected or ``tagged'' particles while $\phi$ is the volume fraction of the granular material, i.e., \emph{all} particles present in a unit area, not just tagged ones.} and $g$ is the acceleration due to gravity.

Unlike molecular solutes \cite{t53,a56} or colloidal suspensions \cite{gs12,ebs77,la87,vsb10}, granular materials are macroscopic and, thus, not subject to thermal fluctuations or ordinary Brownian motion. Nevertheless, inelastic collision between particles can give rise to macroscopic diffusion \cite{hh80,sb76,s93}. The precise theory of diffusion of granular materials is unsettled \cite{cs12} and many models exist. For example, as early as the 1980s, ``shear-induced diffusion'' models were proposed empirically to provide better fits to experimental data \cite{hh80}. In this case, the diffusivity is modeled as $D = D_0(1 + K \dot\gamma)$, for some constants $D_0$ and $K$. Although such an expression can be motivated for hydrodynamically-interacting colloidal particles \cite{gs12}, it appears to be problematic for  granular flows in which if motion ceases ($v_x=0\Rightarrow\dot\gamma=0$) so do the inter-particle collisions, and, hence, we would expect no effective diffusion ($D=0$). 

On the other hand, kinetic theory for hard spheres can be successfully used for dilute granular flows (``granular gases'') \cite{g03}, and it has been suggested that such theories hold (with appropriate corrections) even for a moderately dense volume fraction of $\phi \approx 0.5$ and beyond \cite{at09}. In particular, it has been shown by Savage and Dai \cite{s93,sd93} that
\begin{equation}
	D = \chi(\phi,e) d^2\left|\dot\gamma\right|,
\label{eq:d_gd}
\end{equation}
where $\chi(\phi,e)$ is a dimensionless function that depends solely on the volume fraction $\phi$ and the restitution coefficient $e$ for particle collisions. In this work, we assume that $\phi$ can be taken to be constant to a first approximation in the fully developed steady flow down an incline, hence $\chi=const.$ as well. This assumption is supported by particle-dynamics simulations \cite{sd93}. It should be noted that Eq.~\eqref{eq:d_gd} can also be deduced using only dimensional analysis.

\section{Dispersion in granular shear flow on an incline}

Now, we combine the mathematical results from Section~\ref{sec:math_disp} with the model from Section~\ref{sec:flow_incline}. The Bagnold profile from Eq.~\eqref{eq:Bagnold} can be re-written as
\begin{equation}
v_x(z) = \frac{5}{3}\overline{v_x}\left[1-\left(1-\frac{z}{h}\right)^{3/2}\right],\qquad \overline{v_x} \equiv \frac{2}{5} A h^{3/2}.  
\label{eq:Bagnold2}
\end{equation}
Then, the Savage--Dai diffusivity from Eq.~\eqref{eq:d_gd} becomes
\begin{equation}
D(z) = D_0 \sqrt{1-\frac{z}{h}},\qquad D_0 \equiv \frac{5}{2}\chi d^2\frac{\overline{v_x}}{h}.
\label{eq:DiffBagnold}
\end{equation}
Substituting Eqs.~\eqref{eq:Bagnold2} and \eqref{eq:DiffBagnold} into Eq.~\eqref{eq:d_eff}, we find that
\begin{equation}
\mathcal{D} = \frac{5}{3} \frac{\overline{v_x}}{h} \chi d^2 + \frac{4}{275} \frac{h^3 \overline{v_x}}{\chi d^2} = D_0\left[ \frac{2}{3} + \frac{8}{1375} \left(\frac{h^2}{\chi d^2}\right)^2 \right].
\label{eq:effective_d}
\end{equation}
The inclination angle $\theta$ enters into the effective dispersivity only through the constant $A$ in the mean flow speed $\overline{v_x}$, while the particle diameter enters both the base diffusivity $D_0$ directly and also $\overline{v_x}$ through $A$. Also, note that the effective dispersivity $\mathcal{D}$ depends on the ratio $h/d$ to the fourth power, which can be extremely large given that $d/h \simeq 10^{-5}-1$ in the context of landslides and debris flows, as discussed in the introduction.

By analogy to the fluids context, we can introduce a P\'{e}clet number $Pe = \overline{v_x}h/D_0$ as the ratio of the transverse diffusion and advection time scales. Using the definition of $D_0$ from Eq.~\eqref{eq:DiffBagnold}, $Pe = 2h^2/(5\chi d^2)$, then the effective dispersivity from Eq.~\eqref{eq:effective_d} can be written as $\mathcal{D} = D_0\frac{2}{3}\left( 1 + \frac{3}{55} Pe^2 \right)$. Furthermore, let us introduce the dimensionless variables:\footnote{By the linearity of Eq.~\eqref{eq:cbar}, $c_0$ is arbitrary. For definiteness, it can be taken to be, e.g., $c_0 = \int_{-\infty}^{+\infty} \bar{c}(x,0) \,\mathrm{d} x$ for a finite mass initial condition.} $\bar{c} = c_0\bar{C}$, $t = ({\ell}/\overline{v_x})T$, $z = hZ$, $x = \ell X$, with $h/\ell \equiv \epsilon$, then Eq.~\eqref{eq:cbar} becomes
\begin{equation}
\frac{\partial \bar{C}}{\partial T} + \frac{\partial \bar{C}}{\partial X} = \frac{\epsilon}{Pe}\frac{2}{3} \left(1 + \frac{3}{55} Pe^2 \right)\frac{\partial^2 \bar{C}}{\partial X^2}.
\end{equation}
In dispersion problems, one is typically interested in the release of a finite mass of material, which can be approximated by a point-source initial condition $\bar{C}(X,0) = \delta(X)$, where $\delta(\cdot)$ is the Dirac delta function, subject to decay boundary conditions $\bar{C}(X,T) \to 0$ as $|X|\to \infty$; other initial conditions are possible as well \cite{t53}. 
Switching to the moving frame, where $\xi = (X-T)/\sqrt{3 Pe/(2\epsilon)}$ is the streamwise coordinate, we arrive at the final form of the macrotransport equation:
\begin{equation}
\frac{\partial \bar{C}}{\partial T} = \left(1 + \frac{3}{55} Pe^2 \right)\frac{\partial^2 \bar{C}}{\partial \xi^2}.
\label{eq:mte_gran}
\end{equation}
For the point-source initial condition, the exact solution to the ``dispersion equation'' \eqref{eq:mte_gran} is
\begin{equation}
\bar{C}(\xi,T) = \frac{1}{\sqrt{4\pi \tilde{\mathcal{D}} T}}\exp\left(-\frac{\xi^2}{4 \tilde{\mathcal{D}} T}\right),
\end{equation}
where $\tilde{\mathcal{D}} = 3\mathcal{D}/(2D_0) = 1+(3/55)Pe^2$ using Eq.~\eqref{eq:effective_d}.
In other words, the dispersing material spreads like a Gaussian with diffusivity $\tilde{\mathcal{D}}$ in the moving frame.

Meanwhile, the {classical Taylor--Aris} version of Eq.~\eqref{eq:mte_gran} for plane Couette flow \cite{be93} is
\begin{equation}
\frac{\partial \bar{C}}{\partial T} = \left( 1 + \frac{1}{30} Pe^2 \right)\frac{\partial^2 \bar{C}}{\partial \zeta^2},\qquad \zeta = \frac{X-T}{\sqrt{Pe/\epsilon}}.
\label{eq:mte_couette}
\end{equation}
The effective dispersivities in Eqs.~\eqref{eq:mte_gran} and \eqref{eq:mte_couette} are the same order of magnitude ($3/55 \approx 0.055$, $1/30 \approx 0.033$) for a given $Pe$. Therefore, Taylor--Aris shear dispersion should be an observable phenomenon in rapid dense granular flow, just as it is for molecular solutes in fluids.

\section{Dispersion in a generic 2D shear profile}

More generally, we can consider the shear profiles given by the velocity field
\begin{multline}
v_x(z) = \left(\frac{1+\alpha}{\alpha}\right) \overline{v_x} \left[1 - \left(1 - \frac{z}{h}\right)^\alpha \right] \\ \Rightarrow\quad D = D_0 \left(1 - \frac{z}{h}\right)^{\alpha-1}, \quad D_0 \equiv (1+\alpha)\frac{\overline{v_x}}{h}\chi d^2.
\label{eq:gen_shear}
\end{multline}
For monodisperse materials, we expect that $1\le \alpha \le 2$, where $\alpha = 1$ and $\alpha = 2$ correspond to Couette and Poiseuille flow, respectively, of a Newtonian fluid between two parallel plates, while $\alpha = 3/2$ is the Bagnold profile for granular flow on an incline. For $\alpha < 1$, the velocity profile is convex; such profiles have been measured experimentally \cite{waecmmga11,fsuol14} in bidisperse chute flows, in which significant size segregation occurs.

\begin{figure}[h]
	\centerline{\includegraphics[width=0.8\columnwidth]{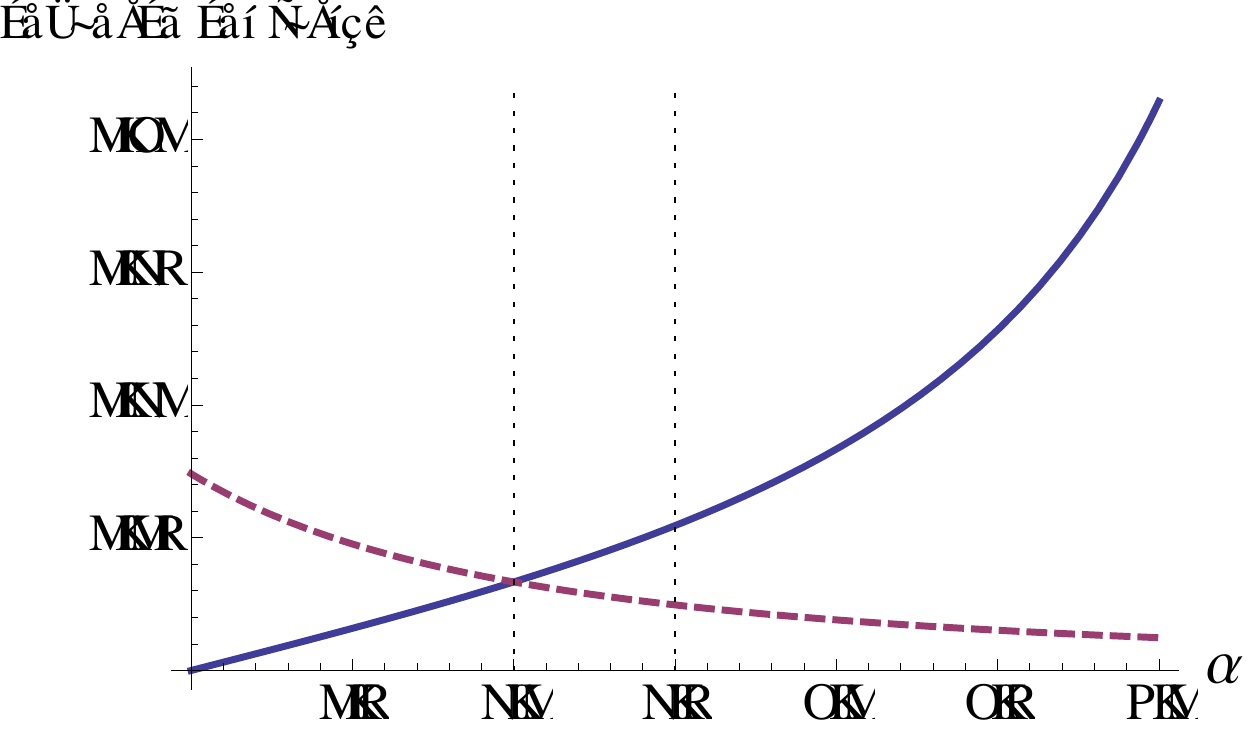}}
	\caption{Enhancement factors (i.e., coefficients of $Pe^2$ in Eq.~\eqref{eq:d_eff_gen_shear}) as functions of the shear profile exponent $\alpha$ in Eq.~\eqref{eq:gen_shear}. The solid curve represents the case of shear-rate-dependent diffusivity, while the dashed curve corresponds to the case of constant diffusivity. Vertical dotted lines are a guide-to-eye representing $\alpha = 1$ (plane Couette flow of a Newtonian fluid) and $\alpha = 3/2$ (Bagnold profile for a dense granular flow down an incline).}
\label{fig:efac}	
\end{figure}

Following the same procedure as above, we obtain the effective dispersivities for such flow profiles:
\begin{equation}
\frac{\mathcal{D}}{D_0} = \begin{cases} \displaystyle\frac{1}{\alpha} \left[1 + \displaystyle\frac{\alpha}{2(4-\alpha)(4+\alpha)} Pe^2\right], \;&D \propto \dot\gamma,\\[4mm] 1 + \displaystyle\frac{2}{3(9+9\alpha+2\alpha^2)} Pe^2, &D = const.,\end{cases}
\label{eq:d_eff_gen_shear}
\end{equation}
with the P\'eclet number defined as before. Let us the define the \emph{enhancement factor} as the coefficient of $Pe^2$ in the expressions in Eq.~\eqref{eq:d_eff_gen_shear}. Figure~\ref{fig:efac} shows the dependence of the enhancement factors on the shear profile exponent $\alpha$. It is evident that for larger $\alpha$, the dispersivity of a material with shear-rate-dependent diffusivity increases significantly over the constant-diffusivity case.

\section{Conclusion}

In this paper, we presented the calculation of the Taylor--Aris effective dispersivity for the rapid flow of a dry, cohesionless monodisperse granular material down an incline, assuming that  volume fraction variations are negligible in the fully-developed Bagnold profile and that the diffusivity is proportional to the shear rate. In particular, for this prototypical granular flow, we found that the enhancement of the diffusivity due to the shear flow varies as the P\'eclet number squared, which is the same dependence found for molecular solutes with constant diffusivity in a shear flow of a Newtonian fluid. This result suggests that shear dispersion is a relevant transport mechanism in flows of granular materials. Moreover, we showed that with increasing concavity of the shear profile, the enhancement factor for a shear-rate-dependent diffusivity grows significantly, while the constant-diffusivity enhancement factor decays. This feature could suggest approaches for maximizing/minimizing dispersion in flows of particulate materials by controlling the shear profile.

A limitation of the present work is that we have assumed, to a first approximation, a constant volume fraction and that the particle flux $\vec{q}$ relative to the flow profile is Fickian, namely $\vec{q} \propto -D\nabla c$, where $D$ is allowed to depend on any of the independent variables, explicitly or implicitly.
Thus, an avenue of future work is to incorporate non-Fickian effects such as volume-fraction variation and segregation of bidisperse materials by generalizing Eq.~\eqref{eq:adv_diff} using mixture theory \cite{gt05}, which leads to the addition of, e.g., a term proportional to $S\dot\gamma\phi(1-\phi)$ in $\vec{q}$, where $S\dot\gamma$ is a percolation velocity (see, e.g., \cite{sl88,waecmmga11,fsuol14}). For the case of granular materials immersed in a viscous fluid (e.g., concentrated colloidal suspensions), shear-induced migration effects due to hydrodynamic interactions \cite{ebs77,la87,vsb10,pabga92} could also be included along these lines by augmenting $\vec{q}$ with a term proportional to $d^2 \phi \nabla(\phi\dot\gamma)$. These extensions of the problem lead to concentration-dependence effects and, consequently, to \emph{nonlinear} dispersion equations (see, e.g., \cite{gs12,yzpb11,gc12}) and/or dispersion processes with streamwise variations of the mean flow speed \cite{sb99}. Finally, in the related context of porous media, it has been suggested that even \emph{nonlocal} effects can arise in the macrotransport equation \cite{kb87} (see also the discussion in \cite{yj91}).

In conclusion, we hope that this work will stimulate further research on the interaction between shear, diffusion and dispersion in flows of granular materials. In particular, it would be of interest to design experiments that lead to the verification of the theoretical results presented herein.

\raggedbottom


\begin{acknowledgements}
I.C.C.\ was supported by the National Science Foundation (NSF) under Grant No.\ DMS-1104047 (at Princeton University) and by the LANL/LDRD Program through a Feynman Distinguished Fellowship (at Los Alamos National Laboratory). LANL is operated by Los Alamos National Security, L.L.C. for the National Nuclear Security Administration of the U.S. Department of Energy under Contract No. DE-AC52-06NA25396.
H.A.S.\ thanks the NSF for support via Grant No.\ CBET-1234500.
We acknowledge useful discussions with Ian Griffiths and Gregory Rubinstein on the derivation of the dispersion equations for the case of non-constant diffusivity, and we thank Ben Glasser for helpful conversations.
\end{acknowledgements}


\appendix
\section*{Appendix}

Following \cite{t53,gs12}, first we substitute $c(x,z,t) \equiv \bar{c}(x,t) + c'(x,z,t)$ and $v_x(z) \equiv \overline{v_x} + v_x'(z)$ into Eq.~\eqref{eq:adv_diff} to obtain%
\begin{multline}
\frac{\partial \bar{c}}{\partial t} + \frac{\partial c'}{\partial t} + \overline{v_x}\frac{\partial \bar{c}}{\partial x} + v_x'\frac{\partial \bar{c}}{\partial x} + \overline{v_x}\frac{\partial c'}{\partial x}  + v_x'\frac{\partial c'}{\partial x} \\
= \frac{\partial}{\partial x}\left(D\frac{\partial \bar{c}}{\partial x}\right) + \frac{\partial}{\partial x}\left(D\frac{\partial c'}{\partial x}\right) + \underbrace{\frac{\partial}{\partial z}\left(D\frac{\partial \bar{c}}{\partial z}\right)}_{=0} + \frac{\partial}{\partial z}\left(D\frac{\partial c'}{\partial z}\right).
\label{eq:adv_diff_split}
\end{multline}
Next, we apply the depth-averaging operator $\overline{(\cdot)} = \frac{1}{h}\int_0^h (\cdot) \,\mathrm{d}z$ to Eq.~\eqref{eq:adv_diff_split} to obtain the governing equation for the depth-averaged concentration:%
\begin{equation}
\underline{\underline{\frac{\mathfrak{D} \bar{c}}{\mathfrak{D} t} + \overline{v_x'\frac{\partial c'}{\partial x}}}}
= \underline{\underline{\frac{\partial}{\partial x}\left(\overline{D}\frac{\partial \bar{c}}{\partial x}\right)}} + \frac{\partial}{\partial x}\left(\overline{D\frac{\partial c'}{\partial x}}\right),\quad \frac{\mathfrak{D}}{\mathfrak{D} t} \equiv \frac{\partial}{\partial t} + \overline{v_x}\frac{\partial}{\partial x},
\label{eq:adv_diff_mean}
\end{equation}
where the average of the last term on the right-hand side of Eq.~\eqref{eq:adv_diff_split} vanishes due to the no-flux boundary condition $\partial c/\partial z = 0$ ($\Rightarrow \partial c'/\partial z = 0$) at $z=0,h$. In Eq.~\eqref{eq:adv_diff_mean} and below, the double-underlined terms turn out to be the dominant ones in the dispersion regime. Now, we subtract Eq.~\eqref{eq:adv_diff_mean} from Eq.~\eqref{eq:adv_diff_split} to obtain the governing equation for the concentration fluctuations:
\begin{multline}
\frac{\mathfrak{D} c'}{\mathfrak{D} t} + \underline{\underline{v_x'\frac{\partial \bar{c}}{\partial x}}} + v_x'\frac{\partial c'}{\partial x} - \overline{ v_x'\frac{\partial c'}{\partial x}}
= \frac{\partial}{\partial x}\left[(D-\overline{D})\frac{\partial \bar{c}}{\partial x}\right]\\ + \frac{\partial}{\partial x}\left(D\frac{\partial c'}{\partial x}\right) + \underline{\underline{\frac{\partial}{\partial z}\left(D\frac{\partial c'}{\partial z}\right)}} - \frac{\partial}{\partial x}\left(\overline{D\frac{\partial c'}{\partial x}}\right).
\label{eq:adv_diff_fluct}
\end{multline}

At this point, we invoke the asymptotic assumptions in the dispersion regime, namely that $|c'| \ll \bar{c}$ once transverse diffusion has equilibrated, i.e., for $\ell/h\gg \overline{v_x}h/D_0$. Meanwhile, both $\overline{v_x}$ and $v_x'$ are the same order of magnitude because the velocity field is steady and given. Thus, the scales for the various variables are
\begin{multline}
[\bar{c}] = c_0,\quad [c'] = \epsilon c_0,\quad [\overline{v_x}] = [v_x'] = U,\quad [D] = D_0,\\ [x] = \ell,\quad [z] = h = \epsilon \ell,
\end{multline}
where $0<\epsilon\ll1$, and the scaling for $z$ is set by the assumption $\ell/h \gg \overline{v_x}h/D_0$, which implies that $h \ll \ell [D_0/(\overline{v_x} h)]$, where $D_0/(\overline{v_x} h)$ is the inverse of the (dimensionless) P\'eclet number, which is assumed to be $\mathcal{O}(1)$.

Now, to ensure that the dispersion problem is nontrivial, both the material derivative and the fluctuation term on the left-hand side of Eq.~\eqref{eq:adv_diff_mean} should be retained, which sets the timescale to be $[t] = \ell/(\epsilon U)$, i.e., we are considering the ``long time'' behavior as posited by Taylor \cite{t53}. Then, upon dividing both sides of Eq.~\eqref{eq:adv_diff_mean} by $\epsilon c_0 U/\ell$ and defining $U h/D_0 = \mathcal{O}(1)$ as the P\'eclet number, it is evident that the first term on the right-hand side of Eq.~\eqref{eq:adv_diff_mean} (underlined) is $\mathcal{O}(1)$, while the second term is $\mathcal{O}(\epsilon)$. Thus, in the dispersion regime, the evolution equation \eqref{eq:adv_diff_mean} of the depth-averaged concentration reduces to Eq.~\eqref{eq:cbar}.

Turning to the left-hand side of Eq.~\eqref{eq:adv_diff_fluct}, we first divide both sides by $c_0 U/\ell$. Then, it is clear only the second term on the left-hand side (underlined) is $\mathcal{O}(1)$, while all other terms are $\mathcal{O}(\epsilon)$ or smaller. Meanwhile, on the right-hand side of Eq.~\eqref{eq:adv_diff_fluct}, again defining $U h/D_0 = \mathcal{O}(1)$ as the P\'eclet number, only the second-to-last term (underlined) is $\mathcal{O}(1)$, while all other terms are $\mathcal{O}(\epsilon)$ or smaller. Thus, in the dispersion regime, the evolution equation \eqref{eq:adv_diff_fluct} of the concentration fluctuations reduces to Eq.~\eqref{eq:cprime}.

Finally, we note that Eq.~\eqref{eq:cbar} and \eqref{eq:cprime} can also be derived formally by perturbation techniques such as the method of multiple time scales \cite{pnb86,man96} with the aspect ratio $\epsilon \equiv h/\ell$ as the small parameter.


\bibliographystyle{unsrt}
\bibliography{granular_diffusion_model}   

\begin{thebibliography}{10}

\bibitem{be93}
H.~Brenner and D.~A. Edwards.
\newblock {\em Macrotransport Processes}.
\newblock Butterworth-Heinemann, Boston, 1993.

\bibitem{t53}
G.~Taylor.
\newblock Dispersion of soluble matter in solvent flowing slowly through a
  tube.
\newblock {\em Proc. R. Soc. Lond. A}, 219:186--203, 1953.

\bibitem{a56}
R.~Aris.
\newblock On the dispersion of a solute in a fluid flowing through a tube.
\newblock {\em Proc. R. Soc. Lond. A}, 235:67--77, 1956.

\bibitem{yj91}
W.~R. Young and S.~Jones.
\newblock Shear dispersion.
\newblock {\em Phys. Fluids A}, 3:1087--1101, 1991.

\bibitem{nht95}
V.~V.~R. Natarajan, M.~L. Hunt, and E.~D. Taylor.
\newblock Local measurements of velocity fluctuations and diffusion
  coefficients for a granular material flow.
\newblock {\em J. Fluid Mech.}, 304:1--25, 1995.

\bibitem{c97}
C.~S. Campbell.
\newblock Self-diffusion in granular shear flows.
\newblock {\em J. Fluid Mech.}, 348:85--101, 1997.

\bibitem{s84}
S.~B. Savage.
\newblock The mechanics of rapid granular flows.
\newblock {\em Adv. Appl. Mech.}, 24:289--366, 1984.

\bibitem{jnb96}
H.~M. Jaeger, S.~R. Nagel, and R.~P. Behringer.
\newblock Granular solids, liquids, and gases.
\newblock {\em Rev. Mod. Phys.}, 68:1259--1273, 1996.

\bibitem{afp13}
B.~Andreotti, Y.~Forterre, and O.~Pouliquen.
\newblock {\em Granular Media: Between Fluid and Solid}.
\newblock Cambridge University Press, Cambridge, 2013.

\bibitem{at09}
I.~S. Aranson and L.~S. Tsimring.
\newblock {\em Granular Patterns}.
\newblock Oxford University Press, New York, 2009.

\bibitem{hd08}
A.~Hacina and D.~Kamel.
\newblock Indirect method of measuring dispersion coefficients for granular
  flow in a column of dihedrons.
\newblock {\em Int. J. Food Eng.}, 4:10, 2008.

\bibitem{setal08}
E.~Simsek, S.~Wirtz, V.~Scherer, H.~Kruggel-Emden, R.~Grochowski, and
  P.~Walzel.
\newblock An experimental and numerical study of transversal dispersion of
  granular material on a vibrating conveyor.
\newblock {\em Particle Sci. Tech.}, 26:177--196, 2008.

\bibitem{i97}
R.~M. Iverson.
\newblock The physics of debris flows.
\newblock {\em Rev. Geophys.}, 35:245--296, 1997.

\bibitem{ph07}
S.~P. Pudasaini and K.~Hutter.
\newblock {\em Avalanche Dynamics}.
\newblock Springer-Verlag, Berlin/Heidelberg, 2007.

\bibitem{gk10}
J.~M. N.~T. Gray and B.~P. Kokelaar.
\newblock Large particle segregation, transport and accumulation in granular
  free-surface flows.
\newblock {\em J. Fluid Mech.}, 652:105--137, 2010.

\bibitem{pds97}
O.~Pouliquen, J.~Delour, and S.~B. Savage.
\newblock Fingering in granular flows.
\newblock {\em Nature}, 386:816--817, 1997.

\bibitem{nist93}
T.~Nakashizuka, S.~Iida, W.~Suzuki, and T.~Tanimoto.
\newblock Seed dispersal and vegetation development on a debris avalanche on
  the {Ontake} volcano, {Central Japan}.
\newblock {\em J. Veget. Sci.}, 4:537--542, 1993.

\bibitem{hh80}
C.~L. Hwang and R.~Hogg.
\newblock Diffusive mixing in flowing powders.
\newblock {\em Powder Technol.}, 26:93--101, 1980.

\bibitem{sl88}
S.~B. Savage and C.~K.~K. Lun.
\newblock Particle size segregation in inclined chute flow of dry cohesionless
  granular solids.
\newblock {\em J. Fluid Mech.}, 189:311--335, 1988.

\bibitem{gs12}
I.~M. Griffiths and H.~A. Stone.
\newblock Axial dispersion via shear-enhanced diffusion in colloidal
  suspensions.
\newblock {\em EPL}, 97:58005, 2012.

\bibitem{jfp06}
P.~Jop, Y.~Forterre, and O.~Pouliquen.
\newblock A constitutive law for dense granular flows.
\newblock {\em Nature}, 441:727--730, 2006.

\bibitem{k11}
D.~V. Khakhar.
\newblock Rheology and mixing of granular materials.
\newblock {\em Macromol. Mater. Eng.}, 296:278--289, 2011.

\bibitem{bdlb09}
D.~Bolster, M.~Dentz, and T.~{Le Borgne}.
\newblock Solute dispersion in channels with periodically varying apertures.
\newblock {\em Phys. Fluids}, 21:056601, 2009.

\bibitem{b54}
R.~A. Bagnold.
\newblock Experiments on a gravity-free dispersion of large solid spheres in a
  {Newtonian} fluid under shear.
\newblock {\em Proc. R. Soc. Lond. A}, 225:49--63, 1954.

\bibitem{seghlp01}
L.~E. Silbert, D.~{Erta\c{s}}, G.~S. Grest, T.~C. Halsey, D.~Levine, and S.~J.
  Plimpton.
\newblock Granular flow down an inclined plane: {Bagnold} scaling and rheology.
\newblock {\em Phys. Rev. E}, 64:051302, 2001.

\bibitem{ebs77}
E.~C. Eckstein, D.~G. Bailey, and A.~H. Shapiro.
\newblock Self-diffusion of particles in shear flow of a suspension.
\newblock {\em J. Fluid Mech.}, 79:191--208, 1977.

\bibitem{la87}
D.~Leighton and A.~Acrivos.
\newblock The shear-induced migration of particles in concentrated suspensions.
\newblock {\em J. Fluid Mech.}, 181:415--439, 1987.

\bibitem{vsb10}
H.~M. Vollebregt, R.~G.~M. {van der Sman}, and R.~M. Boom.
\newblock Suspension flow modelling in particle migration and microfiltration.
\newblock {\em Soft Matter}, 6:6052--6064, 2010.

\bibitem{sb76}
A.~M. Scott and J.~Bridgwater.
\newblock Self-diffusion of spherical particles in a simple shear apparatus.
\newblock {\em Powder Technol.}, 14:177--183, 1976.

\bibitem{s93}
S.~B. Savage.
\newblock Disorder, diffusion, and structure formation in granular flow.
\newblock In A.~Hansen and D.~Bideau, editors, {\em Disorder and Granular
  Media}, pages 255--285. Elsevier, Amsterdam, 1993.

\bibitem{cs12}
I.~C. Christov and H.~A. Stone.
\newblock Resolving a paradox of anomalous scalings in the diffusion of
  granular materials.
\newblock {\em Proc. Natl Acad. Sci. USA}, 109:16012--16017, 2012.

\bibitem{g03}
I.~Goldhirsch.
\newblock Rapid granular flows.
\newblock {\em Annu. Rev. Fluid Mech.}, 35:267--293, 2003.

\bibitem{sd93}
S.~B. Savage and R.~Dai.
\newblock Studies of granular shear flows: Wall slip velocities, ÔlayeringÕ and
  self-diffusion.
\newblock {\em Mech. Mat.}, 16:225--238, 1993.

\bibitem{waecmmga11}
S.~Wiederseiner, N.~Andreini, G.~{\'{E}pely-Chauvin}, G.~Moser, M.~Monnereau,
  J.~M. N.~T. Gray, and C.~Ancey.
\newblock Experimental investigation into segregating granular flows down
  chutes.
\newblock {\em Phys. Fluids}, 23:013301, 2011.

\bibitem{fsuol14}
Y.~Fan, C.~P. Schlick, P.~B. Umbanhowar, J.~M. Ottino, and R.~M. Lueptow.
\newblock Modeling size segregation of granular materials: the roles of
  segregation, advection, and diffusion.
\newblock {\em J. Fluid Mech.}, 714:252--279, 2014.

\bibitem{gt05}
J.~M. N.~T. Gray and A.~R. Thornton.
\newblock A theory for particle size segregation in shallow granular
  free-surface flows.
\newblock {\em Proc. R. Soc. A}, 461:1447--1473, 2005.

\bibitem{pabga92}
R.~J. Phillips, R.~C. Armstrong, R.~A. Brown, A.~L. Graham, and J.~R. Abbott.
\newblock A constitutive equation for concentrated suspensions that accounts
  for shear-induced particle migration.
\newblock {\em Phys. Fluids A}, 4:30--40, 1992.

\bibitem{yzpb11}
A.~Yaroshchuk, E.~Zholkovskiy, S.~Pogodin, and V.~Baulin.
\newblock Coupled concentration polarization and electroosmotic circulation
  near micro/nanointerfaces: {Taylor--Aris} model of hydrodynamic dispersion
  and limits of its applicability.
\newblock {\em Langmuir}, 27:11710--11721, 2011.

\bibitem{gc12}
S.~Ghosal and Z.~Chen.
\newblock Electromigration dispersion in a capillary in the presence of
  electro-osmotic flow.
\newblock {\em J. Fluid Mech.}, 697:436--454, 2012.

\bibitem{sb99}
H.~A. Stone and H.~Brenner.
\newblock Dispersion in flows with streamwise variations of mean velocity:
  Radial flow.
\newblock {\em Ind. Eng. Chem. Res.}, 38:851--854, 1999.

\bibitem{kb87}
D.~L. Koch and J.~F. Brady.
\newblock A non-local description of advection-diffusion with application to
  dispersion in porous media.
\newblock {\em J. Fluid Mech.}, 180:387--403, 1987.

\bibitem{pnb86}
M.~Pagitsas, A.~Nadim, and H.~Brenner.
\newblock Multiple time scale analysis of macrotransport processes.
\newblock {\em Physica A}, 135:533--550, 1986.

\bibitem{man96}
C.~C. Mei, J.-L. Auriault, and C.-O. Ng.
\newblock Some applications of the homogenization theory.
\newblock {\em Adv. Appl. Mech.}, 32:277--348, 1996.

\end{thebibliography}

\end{document}